\documentclass{article}
\usepackage{arxiv}

\usepackage{amsmath}
\usepackage{color}
\usepackage[utf8]{inputenc} 
\usepackage[T1]{fontenc}    
\usepackage{hyperref}       
\usepackage{url}            
\usepackage{booktabs}       
\usepackage{amsfonts}       
\usepackage{nicefrac}      
\usepackage{microtype}      
\usepackage{cleveref}      
\usepackage{lipsum}         
\usepackage{graphicx}
\usepackage{natbib}
\usepackage{doi}
\usepackage[ruled,linesnumbered]{algorithm2e}
\usepackage{bm}
\usepackage{mathrsfs}
\usepackage{listings}

\SetCommentSty{mycommfont}

\SetKwInput{KwInput}{Input}                
\SetKwInput{KwOutput}{Output}

\definecolor{dkgreen}{rgb}{0,0.6,0}
\definecolor{gray}{rgb}{0.5,0.5,0.5}
\definecolor{mauve}{rgb}{0.58,0,0.82}

\title{A Quantum Optimization Method for Geometric Constrained Image Segmentation}

\date{}

\newif\ifuniqueAffiliation
\uniqueAffiliationtrue

\ifuniqueAffiliation 
\author{ 
	\href{https://orcid.org/0000-0003-2556-7612}{
		\includegraphics[scale=0.06]{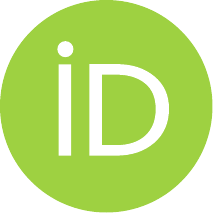} 
		\hspace{1mm}Nam H.~Le}
		\thanks{Corresponding author, alternative email address namhle.vt@gmail.com.} \\
	Department of Electrical and Computer Engineering\\
	The University of Iowa\\
	Iowa City, IA 52242 \\
	\texttt{nam-h-le@uiowa.edu} \\
	\And
	\href{https://orcid.org/0000-0002-9613-9968}{\includegraphics[scale=0.06]{orcid.pdf}\hspace{1mm}Milan Sonka} \\
	Department of Electrical and Computer Engineering\\
	The University of Iowa\\
	Iowa City, IA 52242 \\
	\texttt{milan-sonka@uiowa.edu} \\
	\AND
	\href{https://orcid.org/0000-0003-1372-983X}{\includegraphics[scale=0.06]{orcid.pdf}\hspace{1mm}Fatima Toor} \\
	Department of Electrical and Computer Engineering\\
	The University of Iowa\\
	Iowa City, IA 52242 \\
	\texttt{fatima-toor@uiowa.edu} \\
}
\else
\usepackage{authblk}

\newbox{\orcid}\sbox{\orcid}{\includegraphics[scale=0.06]{orcid.pdf}} 
\author[1]{%
	\href{https://orcid.org/0000-0000-0000-0000}{\usebox{\orcid}\hspace{1mm}David S.~Hippocampus\thanks{\texttt{hippo@cs.cranberry-lemon.edu}}}%
}
\author[1,2]{%
	\href{https://orcid.org/0000-0000-0000-0000}{\usebox{\orcid}\hspace{1mm}Elias D.~Striatum\thanks{\texttt{stariate@ee.mount-sheikh.edu}}}%
}
\affil[1]{Department of Computer Science, Cranberry-Lemon University, Pittsburgh, PA 15213}
\affil[2]{Department of Electrical Engineering, Mount-Sheikh University, Santa Narimana, Levand}
\fi

\renewcommand{\headeright}{Le et al.}
\renewcommand{\undertitle}{}
\renewcommand{\shorttitle}{}

\hypersetup{
pdftitle={A Quantum Optimization Method for Geometric Constrained Image Segmentation}, 
pdfsubject={q-bio.NC, q-bio.QM},
pdfauthor={Nam H.~Le, Milan Sonka, Fatima Toor},
pdfkeywords={quantum computing, quantum algorithm, combinatorial optimization, image segmentation, graph theory, QAOA},
}

\begin{document}
\maketitle

\begin{abstract}
	Quantum image processing is a growing field attracting attention from both the quantum computing and image processing communities. We propose a novel method in combining a graph-theoretic approach for optimal surface segmentation and hybrid quantum-classical optimization of the problem-directed graph. The surface segmentation is modeled classically as a graph partitioning problem in which a smoothness constraint is imposed to control surface variation for realistic segmentation. Specifically, segmentation refers to a source set identified by a minimum s-t cut that divides graph nodes into the source (s) and sink (t) sets. The resulting surface consists of graph nodes located on the boundary between the source and the sink. Characteristics of the problem-specific graph, including its directed edges, connectivity, and edge capacities, are embedded in a quadratic objective function whose minimum value corresponds to the ground state energy of an equivalent Ising Hamiltonian. This work explores the use of quantum processors in image segmentation problems, which has important applications in medical image analysis. Here, we present a theoretical basis for the quantum implementation of LOGISMOS and the results of a simulation study on simple images. Quantum Approximate Optimization Algorithm (QAOA) approach was utilized to conduct two simulation studies whose objective was to determine the ground state energies and identify bitstring solutions that encode the optimal segmentation of objective functions. The objective function encodes tasks associated with surface segmentation in 2-D and 3-D images while incorporating a smoothness constraint. In this work, we demonstrate that the proposed approach can solve the geometric-constrained surface segmentation problem optimally with the capability of locating multiple minimum points corresponding to the globally minimal solution.
\end{abstract}



\section{Introduction}
Previously, an algorithmic framework LOGISMOS, Layered Optimal Graph Image Segmentation of Multiple Objects and Surfaces (\cite{liGloballyOptimalSegmentation2004,liOptimalSurfaceSegmentation2006,wuOptimalNetSurface2002,zhangChapter11LOGISMOSJEI2020}), was developed and its effectiveness in segmentation of multiple interacting surfaces have been proven in clinical applications (\cite{kashyapJustEnoughInteraction2017,oguzLOGISMOSBLayeredOptimal2014, leSemiautomatedIntracranialAneurysm2022}).
The principle of LOGISMOS lies in the reformulation of the surface optimization task to the problem of finding a minimum $s-t$ cut in a directed graph. This representation is flexible because geometric constraints can be easily incorporated into the graph construction by additions of graph edges with infinite capacities.

Krauss and colleagues (\cite{kraussSolvingMaxFlowProblem2020}) presented three methods to formulate the maximal flow as a quadratic unconstrained binary optimization (QUBO) problem. In a QUBO formulation, the objective function can be expressed as a quadratic polynomial of binary variables. The cost matrix can be converted to a Hamiltonian matrix implemented by a series of rotation Z gates. The problem of finding a minimum function value of QUBO is equivalent to finding the ground state energy of the corresponding problem Hamiltonian.

Estimating the ground state of a random Hamiltonian is difficult. The Quantum Approximate Optimization Algorithm (QAOA) (\cite{farhiQuantumApproximateOptimization2014}) provides an adiabatic way to evolve a ground state of a simple mixing Hamiltonian to a complete problem Hamiltonian. QAOA is a special case of variational quantum circuits. At each layer of QAOA, a pair of operators called cost and mixer unitaries are applied to slowly perturb the system Hamiltonian to the target Hamiltonian. In order to minimize the expectation value of the problem Hamiltonian, the parameters of the unitaries are subjected to classical optimization. This optimization process is iteratively repeated until no further improvements in the minimum objective function value can be achieved.

Through our analysis, we have discovered that the minimum cut formulation when optimized by QAOA provides a stochastic optimization approach for LOGISMOS surface segmentation which can return valid solutions at different runs.

\section{Method}
\label{sec:headings}

\begin{figure}[htbp]
    \centering
    \includegraphics[width=0.8\linewidth]{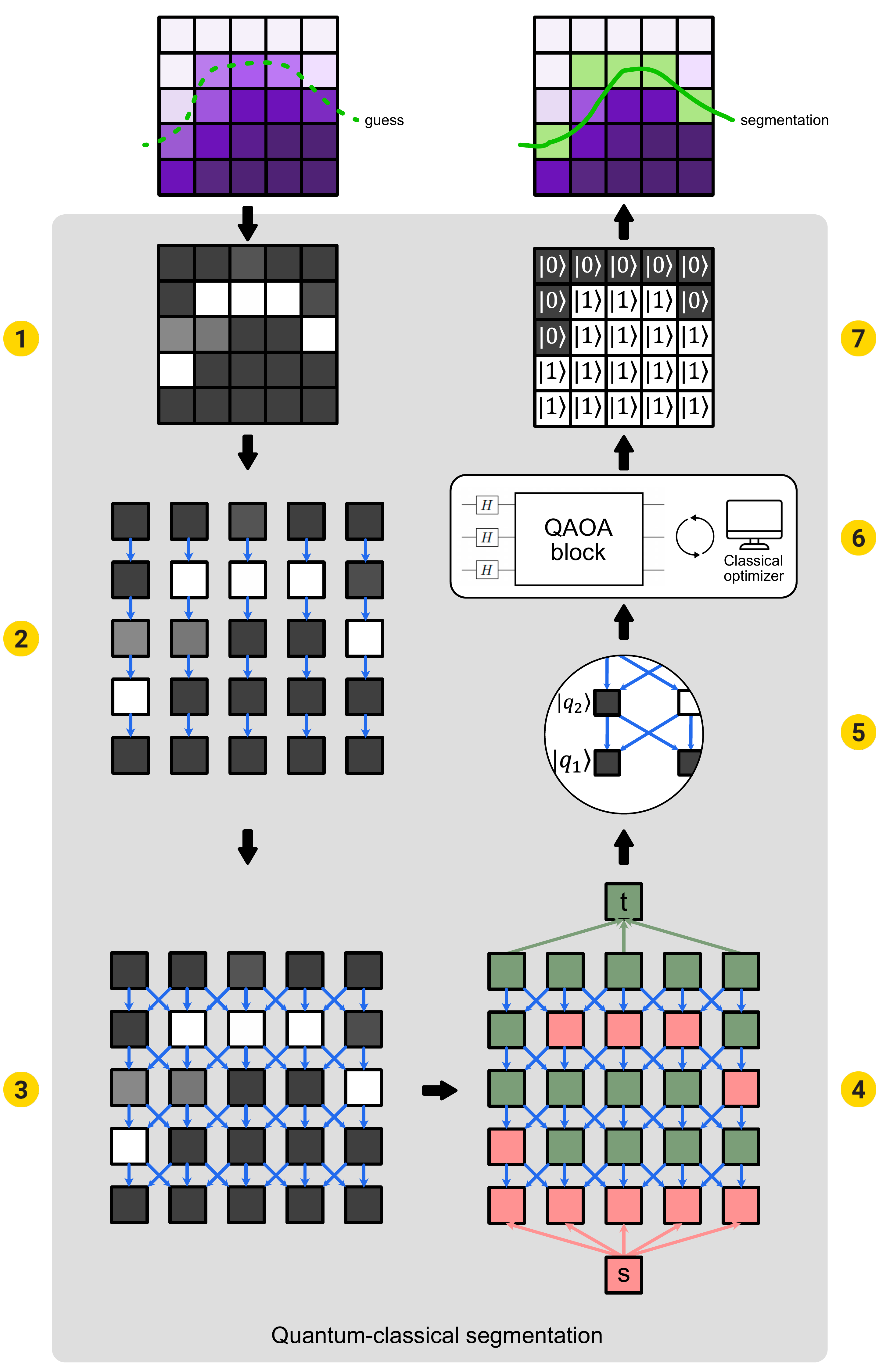}
    \caption{The proposed Quantum LOGISMOS framework: (1) Estimate cost functions and calculate terminal weights,  (2) Introduce internal connections within columns to ensure that the optimal surface cut passes through each column only once, (3) Add inter-edges to impose smoothness condition, (4) Add source and sink nodes, add problem-specific edges, (5) Assign qubit to graph node, (6) QAOA optimization, (7) Bitstring solution and minimum closed set found.
    } 
    \label{fig:method}
\end{figure}

The proposed QuantumLOGISMOS framework is depicted in Fig. \ref{fig:method}. The process begins with an input image, and the primary objective is to determine an optimal surface that separates a region of interest from the background. This is achieved by defining a cost function that emphasizes intensity changes within the image. The solution to the segmentation problem lies in identifying the nodes whose associated costs minimize the total cost function. Originally, this task is approached by recognizing that the optimal surface nodes are within a closed set, amounting to a minimal total terminal weight. Such a set can be found by cutting the graph with a minimum $s-t$ cut in a directed graph, whose edge capacities dictate the maximum flow through the network. To initiate the graph, each image pixel is represented as a graph node. We augment the graph with a series of infinite capacity edges that enforce constraints on the permissible variations in smoothness for the solution surface. To determine the minimum $s-t$ cut that separates the segmentation set from the background, we minimize a QUBO objective function using a hybrid quantum-classical QAOA schedule. The resulting bitstring solution represents the optimal surface within the image.

\subsection{LOGISMOS for Single Surface Detection}
\label{sec:logismosforsinglesurfacedetection}
The LOGISMOS framework models a surface segmentation problem as a maximal-flow minimum-cut problem (\cite{dantzigMaxFlowMin1955}). An optimal surface is the boundary of the resulting graph after cutting it along the min $s-t$ edges.

A directed graph $\mathcal{G}=(\mathcal{V},\mathcal{E})$ is constructed from the volumetric image $\mathcal{I}(\mathbf{x}, \mathbf{y}, \mathbf{z})$ satisfying the following properties: (i) each pixel (2-D image) or voxel (3-D image) is represented by a graph node and (ii) all nodes are connected by directed edges to their 4-neighbor or 6-neighbor settings in 2-D or 3-D images, respectively.

A cost function maps graph nodes to cost values which are defined as the unlikeliness of a node to reside on a surface $s$. The lower the cost indicates the greater likeliness of the node to be on the optimal surface. Given a surface function $s(x) = k$ that maps a column $x$ to a node $k$ of a optimal surface $\mathcal{S}$, the corresponding total cost function of the given surface function $s$ is defined as Eq. \eqref{eq:on-surface_cost}

\begin{equation}
    C_s = \sum_{x \in Col_\mathcal{G} } c_s(x, k) = \sum_{x \in Col_\mathcal{G} } c_s(x, s(x)) .
    \label{eq:on-surface_cost}
\end{equation}

Equation \eqref{eq:on-surface_cost} is the sum of the cost of all nodes $k$, one per column in the set of columns $Col_\mathcal{G}$ of the graph $\mathcal{G}$, on the optimal surface $\mathcal{S}$. The optimal surface
$\hat{\mathcal{S}}$ is characterized by the surface function $\hat{s}(x)$ that minimizes the total cost function $C_s$ represented in Eq. \eqref{eq:approx-surface-solution}

\begin{equation}
    \hat{s}(x)=\arg \min _{\hat{s}(x)} C_s .
	\label{eq:approx-surface-solution}
\end{equation}

The problem of searching for an optimal surface can be reformulated as the task of finding a closed set with minimal total terminal weights. This formulation is equivalent to minimizing the total cost function, $\min W_s = \min C_s$. The total terminal weight $W_s$ of a closed set is determined by summing the terminal weights of a subset of nodes $k=1,\ldots,K_x$ belonging to the corresponding column $x$, defined as Eq. \eqref{eq:total-terminal-weight}

\begin{equation}
    W_s = \sum_{x \in Col_\mathcal{G}} \sum_{k^{\prime}=1}^{K_x} w_s\left(x, k^{\prime}\right) ,
	\label{eq:total-terminal-weight}
\end{equation}

\noindent where

\begin{equation}
    w_s(x, k)=
    \begin{cases}
        -1, & \text { if } k=1 \\ c_s(x,k)-c_s(x,k-1), & \text { otherwise } .
    \end{cases}
\end{equation}

Now, we specify how to construct directed edges $\mathcal{E}$ of the graph which characterizes a number of segmentation configurations. First, we construct a set of infinite-capacity directed edges $\mathcal{E}_{\text {intra }}$ imposing convex constraints on the surface $\mathcal{S}$. This ensures each graph column has exactly one cut, in other words, the surface function $s(x) = k$ is guaranteed to be bijective. The smoothness constraint is imposed by adding a series of edges $\mathcal{E}_{\text {inter }}$ among adjacent columns, which enables the realistic delineation of the outlines of anatomical structures and is among the main advantages of the LOGISMOS framework.

Finally, the problem-specific edges imposing the terminal weights are embedded in $\mathcal{E}_{\text {W}}$. To construct $\mathcal{E}_{\text {W}}$, we first add a source node $s$ and a sink $t$ node to the graph $\mathcal{G}=(\mathcal{V},\mathcal{E})$, resulting in $\mathcal{G}_{st}=(\mathcal{V}_{st},\mathcal{E})_{st}$ where $\mathcal{V}_{st} = V \cup \lbrace s,t \rbrace$. For each node $v$ belonging to the set of nodes with negative costs $\mathcal{V}^-$, we add directed edges from the source node $s$ to it with capacity $\lvert w_{s,v} \rvert$. For those nodes belonging to the set $\mathcal{V}^+$ of positive costs, we add directed edges from those to the sink node $t$ with capacity $\lvert w_{v,t} \rvert$. The resulting edges of the LOGISMOS graph $G$ are

\begin{equation}
    \mathcal{E}=\mathcal{E}_{\text {intra}}+\mathcal{E}_{\text {inter}}+\mathcal{E}_{\text {W}} .
\end{equation}

\subsection{Conversion to an Equivalent QUBO}
Building upon the work conducted by Krauss et al. (\cite{kraussSolvingMaxFlowProblem2020}), we develop a QUBO objective function $F_C$ that enables the simultaneous separation of the vertex set $V$ into a source set $s$ and a sink set $t$ when minimized. Here, $F_C$ is a sum of individual objective functions $F_{\langle i, j\rangle}$ for each edge $\langle i, j\rangle \in \mathcal{E}$ and $F_{\langle s, t\rangle}$ for the source-sink edge $\langle s, t\rangle$, as in Eq. \eqref{eq:objective-function-normal-edge} and Eq. \eqref{eq:objective-function-st-edge}

\begin{equation}
    F_{\langle i, j\rangle} = x_i^2-x_i x_j=\left\{\begin{array}{c}
        1, \text { if } i \in \text {source and } j \in \operatorname{sink} \\
        0, \text { otherwise }
    \end{array}\right.
	\label{eq:objective-function-normal-edge}
\end{equation}

\begin{equation}
    F_{\langle s, t\rangle} = -(x_s^2 - x_s x_t)=\left\{\begin{array}{c}
        -1, \text { if cut is valid } \\
        0, \text { otherwise }
    \end{array}\right.
	\label{eq:objective-function-st-edge}
\end{equation}

\noindent which leads to
\begin{equation}
    F_C(x) = \sum_{i, j \in \mathcal{V}+\{s, t\}} \lvert w_{\langle i, j\rangle} \rvert F_{\langle i, j\rangle} + \varepsilon F_{\langle s, t\rangle} .
\end{equation}

The penalty coefficient $\varepsilon$ ensures any feasible cut will result in negative energy.
In our application, we choose the penalty coefficient to be the sum of all capacities plus one, assuming integer terminal weights

\begin{equation}
    \varepsilon = 1 + \sum_{\langle i, j\rangle \in \mathcal{E}} \lvert w_{\langle i, j\rangle} \rvert.
\end{equation}

We are now ready to construct a QUBO objective function. Specifically, the QUBO problem is expressed as

\begin{equation}
    \text{minimize } F_C = \bm{x}^T \bm{Q} \bm{x} = \sum_{i,j}{Q_{ij}x_ix_j}  ,
    \label{eq:qubo_objective_function_general}
\end{equation}

\noindent where $\bm{x}=[x_1, x_2, \ldots]$ is a vector whose elements are the encoded variables specific for the optimization problem and $x_i$ are binary and subjected to $x_i\in \lbrace 0,1 \rbrace$ (\cite{gloverTutorialFormulatingUsing2019}).

Given a LOGISMOS graph, an adjacent matrix $\bm{A}$ can be determined in a square matric $\bm{A}$ whose matrix elements are given in Eq. \eqref{eq:original-adjacent-matrix}

\begin{equation}
    \bm{A} = \left[
        \begin{matrix}
            a_{11} & a_{12} & \ldots & a_{1j} \\a_{21}&a_{22}&\ldots&\vdots\\\vdots&\vdots&\ddots&\vdots\\a_{i1}&\cdots&\cdots&a_{ij}\\
        \end{matrix}
        \right] .
	\label{eq:original-adjacent-matrix}
\end{equation}

\noindent Since LOGISMOS graphs have no loops or self-edges, the diagonal elements of $\bm{A}$ are zero and $\bm{A}$ becomes to Eq. \eqref{eq:ajacent-matrix-diag}

\begin{equation}
    \bm{A} = \left[
        \begin{matrix}
            0 & a_{12} & \ldots & a_{1j} \\a_{21}&0&\ldots&\vdots\\\vdots&\vdots&\ddots&\vdots\\a_{i1}&\cdots&\cdots&0\\
        \end{matrix}\right] .
	\label{eq:ajacent-matrix-diag}
\end{equation}

\noindent Based on the edge construction rules specified in Section \ref{sec:logismosforsinglesurfacedetection}, the matrix elements of $\bm{A}$ are $a_{ij} = \lvert w_{\langle i, j\rangle} \rvert := w_{ij}$, therefore we get $\bm{A}$ of the form shown in Eq. \eqref{eq:ajacent-matrix-terminalweight}

\begin{equation}
    \bm{A} = \left[
        \begin{matrix}
            0      & w_{12} & \ldots & w_{1j} \\
            w_{21} & 0      & \ldots & \vdots \\
            \vdots & \vdots & \ddots & \vdots \\
            w_{i1} & \cdots & \cdots & 0      \\
        \end{matrix}\right] .
	\label{eq:ajacent-matrix-terminalweight}
\end{equation}

The square matrix of constant $\bm{Q}$ can be derived from $\bm{A}$ by alternating the signs of off-diagonal elements of $\bm{A}$. The diagonal elements of $\bm{Q}$ are obtained by summing all outgoing edge capacities of each node $i$ in $\bm{A}$ as in Eq. \eqref{eq:qubo-q-matrix}

\begin{equation}
    \bm{Q} = \left[\begin{matrix}\sum_{j}w_{1j}&{-w}_{12}&\ldots&{-w}_{1j}\\{-w}_{21}&\sum_{j}w_{2j}&\ldots&\vdots\\\vdots&\vdots&\ddots&\vdots\\{-w}_{i1}&\cdots&\cdots&\sum_{j}w_{ij}\\\end{matrix}\right] .
	\label{eq:qubo-q-matrix}
\end{equation}

\noindent Thus, the objective function $F_C$ can be rewritten in QUBO form as Eq. \eqref{eq:qubo_objective_function_logismos}

\begin{equation}
    F_C = \bm{x}^T\bm{Qx} = \bm{x}^T\left[\begin{matrix}\sum_{j}w_{1j}&{-w}_{12}&\ldots&{-w}_{1j}\\{-w}_{21}&\sum_{j}w_{2j}&\ldots&\vdots\\\vdots&\vdots&\ddots&\vdots\\{-w}_{i1}&\cdots&\cdots&\sum_{j}w_{ij}\\\end{matrix}\right]\bm{x} .
    \label{eq:qubo_objective_function_logismos}
\end{equation}


\subsection{Formulation of the Problem Hamiltonian}

We need to find the equivalent problem Hamiltonian that models $F_C$ such that

\begin{equation}
    H_C \left|\psi\right\rangle = F_C \left|\psi\right\rangle.
    \label{eq:qubo_problem_hamiltonian}
\end{equation}

\noindent Equation \eqref{eq:qubo_problem_hamiltonian} indicates that the ground state energy of a quantum system characterized by the Hamiltonian $H_C$ is equivalent to the minimum value of the objective function $F_C$, achieved with the ground state of the quantum system.
Multiplying both sides of Eq. \eqref{eq:qubo_problem_hamiltonian} by bra $\left\langle\psi\right|$ yields

\begin{equation}
    F_C = \left\langle \psi \right | H_C \left|\psi\right\rangle = \left\langle H_C \right\rangle ,
    \label{eq:qubo_objective_function_expectation}
\end{equation}

\noindent which suggests a way to rewrite the QUBO expression of $F_C$ in Eq. \eqref{eq:qubo_objective_function_logismos} as an expectation value of the problem Hamiltonian $H_C$ in the state $\left|\psi\right\rangle$.

From Eq. \eqref{eq:qubo_objective_function_logismos} and Eq. \eqref{eq:qubo_objective_function_expectation}, we have

\begin{equation}
    F_C = \left\langle \psi \right | H_C \left|\psi\right\rangle = \bm{x}^T\bm{Qx} .
\end{equation}

The problem Hamiltonian $H$ corresponding to the objective function $F_C$ that translate binary $\bm{x}$ from classical variables to quantum representation of qubit states $\left|\psi\right\rangle$ represented by Eq. \eqref{eq:problem-hamiltonian}

\begin{equation}
    H_C=\sum_{i,j}{Q_{ij}\frac{1-Z_i}{2}}\frac{1-Z_j}{2} .
	\label{eq:problem-hamiltonian}
\end{equation}

The state $\left|\psi\right\rangle = \left|\psi_1\right\rangle \otimes \left|\psi_2\right\rangle \otimes ... \otimes \left|\psi_n\right\rangle$, $\psi_n \in\{0,1\}^n$ in the Hilbert space $\left(\mathscr{C}^2\right)^{\otimes n}$ encodes the classical bitstrings $\bm{x} = [x_1, x_2, ..., x_n]$. The connection between the value of the QUBO objective function and the energy of the quantum system is as in Eq. \eqref{eq:quantum-classical-connection}

\begin{equation}
    F_C(\bm{x}) = \left\langle\psi\right| H_C \left|\psi\right\rangle = \left\langle\psi\right| E \left|\psi\right\rangle = \langle H_C \rangle = E .
	\label{eq:quantum-classical-connection}
\end{equation}

The solution to the QUBO problem is a superposition state of all qubits representing graph nodes in a QAOA circuit estimating the problem Hamiltonian $H_C$ (\cite{lucasIsingFormulationsMany2014}), that put the circuit energy to ground state energy $E_0$ of the problem Hamiltonian $H_C$ given by Eq. \eqref{eq:ground-state}

\begin{equation}
    \min F_C = E_0 .
	\label{eq:ground-state}
\end{equation}

\subsection{Hybrid Quantum-Classical Optimization}

Locating the ground state of an arbitrary Hamiltonian is nontrivial.
QAOA, proposed by Farhi et al. (\cite{farhiQuantumApproximateOptimization2014}), aims to approximate the ground state of a given problem Hamiltonian $H_C$. The idea of QAOA is based on the adiabatic theorem which states that a quantum system remains in its specific energy state if the Hamiltonian changes slowly enough. 
The framework consists of three parts: (1) prepare an initial ground state of a known and "easy" Hamiltonian $H_M$, (2) a parameterized quantum circuit that slowly evolves the initial state and ground state of the simple Hamiltonian $H_M$ to the final state and supposedly ground state of the problem Hamiltonian $H_C$, and (3) a classical module to get the average expectation values of all shots and update the circuit parameter for the next optimization iteration. The process is repeated until the classical optimizer is no longer able to improve $\theta$.

A QAOA circuit consists of trotterized unitary operators $U_C(\gamma)$, the time evolution operator imposed by $H_C$, and $U_M(\beta)$, a time evolution operator imposed by a simple Hamiltonian $H_M$. $U_C(\gamma)$ and $U_M(\beta)$ are also called problem and mixer unitaries, represented in Eq. \eqref{eq:problem-unitary} and Eq. \eqref{eq:mixer-unitary}, respectively

\begin{equation}
    U_C(\gamma) = e^{-i \gamma H_C}  ,
	\label{eq:problem-unitary}
\end{equation}

\begin{equation}
    U_M(\beta) = e^{-i \beta H_M}  .
	\label{eq:mixer-unitary}
\end{equation}

These pairs of unitary operators are applied to the initial state $\left| \psi_0 \right\rangle$ $p$ times though the parameters $\gamma_i$ and $\beta_i$ at layer $i$, i.e., $\bm{\theta}_i = (\gamma_i, \beta_i)$, resulting in a final state as in Eq. \eqref{eq:final-state}

\begin{equation}
    | \psi(\bm{\gamma}, \bm{\beta}) \rangle =
    U\left(B, \beta_p\right) U\left(C, \gamma_p\right) \cdots U\left(B, \beta_1\right) U\left(C, \gamma_1 \right)| \psi_0 \rangle
	\label{eq:final-state}
\end{equation}

By the adiabatic principle, the Hamiltonian induced by the trotterized circuit is an approximation of the problem Hamiltonian $H_C$ with deeper $p$

\begin{equation}
    U({\bm\theta}) = \prod_{i=1}^p U_C\left(\gamma_i\right) U_M\left(\beta_i\right)
    \approx e^{-iH_Ct} .
\end{equation}

The depth $p$ of the circuit allows for finer perturbation of the QAOA Hamiltonian, i.e., $\bm{\theta}_i$ varies more slowly at each parameterized block. When $p$ approaches infinity, it is theorized that the QAOA circuit is guaranteed to evolve the initial ground state of $H_M$ to the approximation of the ground state of $H_C$.
The final state $| \psi(\bm{\gamma}, \bm{\beta}) \rangle = | \psi (\bm\theta) \rangle$ of the qubits is roughly estimated in the computational basis after a number of shots at the end of each QAOA execution. The energy of the system is then calculated and represented by Eq. \eqref{eq:qaoa-energy}

\begin{equation}
    E\left( \bm\theta \right) = \left\langle\psi\left(\bm\theta\right)\right| H_C \left|\psi\left(\bm\theta\right)\right\rangle = F_C(\bm{x}) .
	\label{eq:qaoa-energy}
\end{equation}

The process is repeated until the classical optimizer is no longer able to update $\bm{\theta}$. The resulting bitstring $\bm{x}$ corresponding to the computational basis is the approximate solution to the QUBO problem. In other words,

\begin{equation}
    \min_{\bm\theta} E\left(\bm\theta\right)
    \approx \min \langle H_C \rangle ,
\end{equation}

\noindent and

\begin{equation}
    \min_{\bm\theta} E\left(\bm\theta\right)
    \approx \min_{\bm{x}} F_C(\bm{x}) .
\end{equation}

\subsection{QuantumLOGISMOS}

The proposed QuantumLOGISMOS algorithm is presented in \ref{alg:qlogismos}. The algorithm takes an image $\mathcal{I}$ as input and returns a surface $\mathcal{S}$ as output. The algorithm consists of three main stages: (1) construct a LOGISMOS graph from the image $\mathcal{I}$, (2) construct a QUBO objective function $F_C$ from the LOGISMOS graph, and (3) run the QAOA algorithm to find the optimal surface $\mathcal{S}$.

\begin{algorithm}[hp]
    \DontPrintSemicolon

    \KwInput{Image $\mathcal{I}$}
    \KwOutput{Surface $\mathcal{S}$}

    Calculate cost function $c_s(x, k)$ \;
    Calculate terminal weights $w_s(x, k)$ \;

    Contruct edge set $\mathcal{E}_{\text {intra }}$ \;

    Set smoothness constraint $\delta$ \;
    Contruct edge set $\mathcal{E}_{\text {inter }}$ \;

    Add a source node $s$ and a sink $t$ node \;

    Contruct edge set $\mathcal{E}_{\text {W }}$ \;

    Get adjacency matrix $\bm{A}$ \;

    Calculate the QUBO matrix $\bm{Q}$ \;

    Choose QAOA circuit depth $p$

    Assign qubit registers to graph nodes, $q_i := x_i$\;

    $E_{\text{min}} = 0$ \;

    \While{$ E(\bm\theta) < E_{\text{min}} $}
    {
        $E_{\text{min}} = E(\bm\theta)$ \;
        Run QAOA circuit with parameters $\bm\theta$ \;

        Calculate $E(\bm\theta)$ \;
        Update $\bm\theta$ \;
    }

    $ \mathcal{S} $ is the set of highest nodes at each column of $G$ \;

    \caption{QuantumLOGISMOS}
    \label{alg:qlogismos}
\end{algorithm}


\section{Experiments}



\subsection{Python Implementation} 
The code snippet Listing \ref{lst:pythoncode} provides a detailed implementation of the core simulation part of our implementation. First, a LOGISMOS graph was constructed from the provided cost matrix and specified smoothness parameter (lines 8-10). In Line 18, the total capacity is calculated by summing the individual capacities of graph edges. Lines 26 to 29 implement the QUBO formulation of LOGISMOS. The Qiskit quantum simulator was used to run the QAOA simulation on the given QUBO objective function (lines 31 to 40). The number of QAOA runs depends on whether the classical optimizer is capable of finding a new set of parameters $\theta$. Otherwise, the QAOA optimization is terminated.

\begin{lstlisting}[caption={Core Python code for the simulation of the QAOA algorithm for the LOGISMOS problem},label={lst:pythoncode},captionpos=t]
    def solve(self):
        """
        Given a Logismos graph, convert the max-flow min-cut problem to a QUBO problem.

        :param graph: the LOGISMOS graph to solve for the minimum s-t cut
        :param reps: depth of the QAOA circuit
        """
        log_graph = self.graph.graph
        source = self.graph.source
        sink = self.graph.sink

        n = len(log_graph.nodes())
        edges = list(log_graph.edges(data=True))

        # sum all capacities of the edges in the graph
        total_capacity = 0
        for edge in log_graph.edges(data=True):
            total_capacity += edge[2]['capacity']


        model_logismos2d = Model()

        x = model_logismos2d.binary_var_list(n)

        # Define the objective function QUBO to be MINIMIZED
        model_logismos2d.minimize(
            model_logismos2d.sum(
                w.get('capacity') * (x[int(i) - 1] * x[int(i) - 1] - x[int(i) - 1] * x[int(j) - 1]) for i, j, w in edges) +
            (total_capacity + 1) * (-x[source - 1] * x[source - 1] + x[source - 1] * x[sink - 1]))
        
        problem = from_docplex_mp(model_logismos2d)

        seed = 1234
        algorithm_globals.random_seed = seed

        spsa = SPSA(maxiter=250)
        sampler = Sampler()
        qaoa = QAOA(sampler=sampler, optimizer=spsa, reps=self.reps)
        algorithm = MinimumEigenOptimizer(qaoa)
        result = algorithm.solve(problem)

        self.objfunc_value = result.fval
        self.segmentation_set = [qubit_index + 1 for qubit_index, qubit_val in enumerate(result.x) if qubit_val == 1]
        self.background_set = [qubit_index + 1 for qubit_index, qubit_val in enumerate(result.x) if qubit_val == 0]

        solution = ['x%d = %d' % (i + 1, x) for i, x in enumerate(result.x)]

        print('Solution is:', solution)
        print('Objective function value:', result.fval)
        print('Segmentation set:', self.segmentation_set)
        print('Background set:', self.background_set)
\end{lstlisting}

The classical optimizer used is Simultaneous Perturbation Stochastic Approximation (SPSA) optimizer (\cite{spallOVERVIEWSIMULTANEOUSPERTURBATION1998}) implemented in Qiskit-optimization 0.5.0. The maximum number of classical optimizing iterations was set to 250. Mixer layers were not used.

A Python implementation of the proposed QuantumLOGISMOS framework was developed. The code was written in Python 3.10.11. Qiskit 0.43.1 and Networkx 2.8.4 were used to construct directed graphs and conduct quantum simulations.

\subsection{Evaluation}

Solutions given by the QAOA method were compared to the results of the classical highest-label preflow-push algorithm (\cite{goldbergNewApproachMaximumflow1988}). Specifically, the cut produced by QAOA is considered a minimum $s-t$ cut if it (i) completely separates the source from the sink and (ii) achieves a minimum total capacity.


\section{Results and Discussion}
\begin{figure*}[hbtp]
    \centering
    \includegraphics[width=0.9\linewidth]{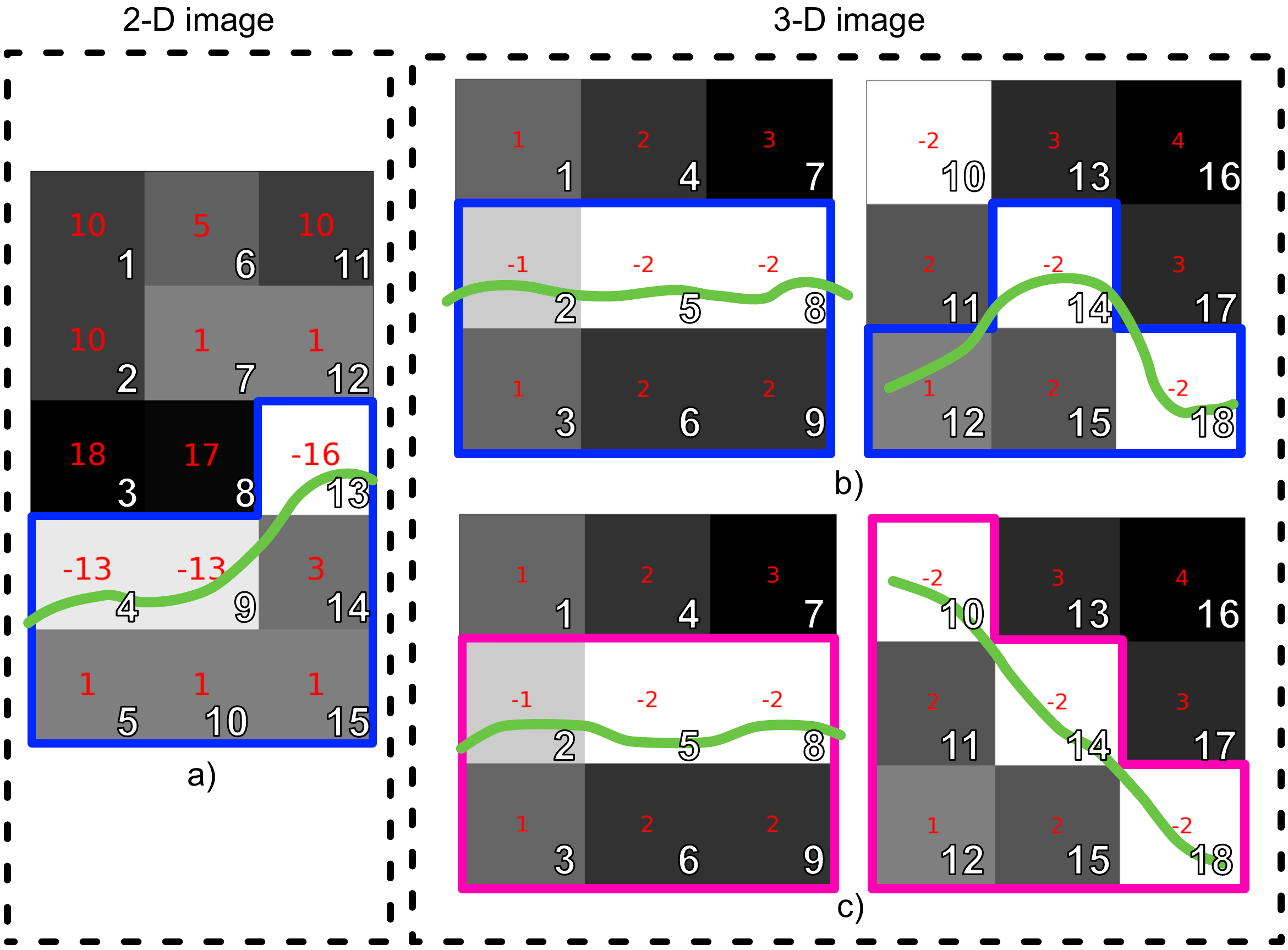}
    \caption{Segmentation results (green lines) given by the proposed method in a simulation study done on an (a) 2-D image sized $3\times5$ and (b-c) 3-D image sized $3\times3\times2$. The terminal weights are colored red and qubit assignments are colored white. For instance, the upper-left corner pixel of the 2-D example has a terminal weight of $10$ and is assigned to qubit $q_1$. The solution to the minimum closed set problem is within the blue and pink boundaries. In the 3-D example, the first row shows one of two optimal segmentation results and the second row shows another optimal segmentation solution.}
    \label{fig:example}
\end{figure*}

Fig. \ref{fig:example} shows two simulation studies, in which the proposed quantum optimization scheme was applied to a 2-D image and a 3-D image. The 2-D image is a $5\times4$ grid of pixels, which were used to construct a graph of three columns with five nodes each. The 3-D image contains two $3\times3$ slices, which were used to construct a graph of three columns with three nodes each. The terminal weights were pre-determined and the optimal surfaces are expected to be along the negatively-weighted nodes. LOGISMOS graphs were constructed with a smoothness constraint $\delta=2$. The quantum optimization was performed on a quantum simulator provided by Qiskit optimization package (\cite{qiskitLectureIntroductionQuantum2021}).

In the first experiment (Fig.\ \ref{fig:example}a), a QAOA solver with five repeated parameterized blocks $p=5$, was able to find a minimum-energy state $\left| q_1 q_2 \ldots q_{15} q_s q_t\right\rangle = \left| 00011000110011110 \right\rangle$ and a ground-state energy $F=-238$. This corresponds to a source set $\{q_4, q_5, q_9, q_{10}, q_{13}, q_{14}, q_{15}, q_{s}\}$ and the optimal surface $\mathcal{S}=\{q_4,q_9,q_{13}\}$ with a flow value $3$.

Fig.\ \ref{fig:example}b and \ref{fig:example}c show two valid optimal surfaces to the second segmentation task found by the quantum-classical approach with varying numbers of repetitions $p=2,3,4,5,6$. In both cases, the maximum flow value was $2$ and the ground-state energy was $F=-162$. The classical highest-label preflow-push algorithm (\cite{goldbergNewApproachMaximumflow1988}), on the other hand, was only able to find the second solution.

\begin{table}[htp]
    \centering
    \caption{Runtime of the classical QAOA simulation based on a varying number of repetitions $p$ of the parameterized blocks.}
    \begin{tabular}{crrrp{10.165em}}
        \toprule
        \multicolumn{1}{p{8.25em}}{Reps of QAOA circuit (complexity)} & \multicolumn{1}{l}{Time (sec)} & \multicolumn{1}{l}{Is solution?} & \multicolumn{1}{p{4.335em}}{Obj func value} & \multicolumn{1}{l}{Solution}                   \\
        \midrule
        2                                                             & 35570                          & TRUE                             & -162                                        & [2,3,5,6,8,9, \newline{}12,14,15,18, 19]       \\
        3                                                             & 27323                          & TRUE                             & -162                                        & [2,3,5,6,8,9, \newline{}12,14,15,18, 19]       \\
        4                                                             & 34752                          & TRUE                             & -162                                        & [2,3,5,6,8,9, \newline{}12,14,15,18, 19]       \\
        5                                                             & 43025                          & TRUE                             & -162                                        & [2,3,5,6,8,9, \newline{}10,11,12,14,15,18, 19] \\
        6                                                             & 29098                          & TRUE                             & -162                                        & [2,3,5,6,8,9, \newline{}12,14,15,18, 19]       \\
        100                                                           & 123901                         & TRUE                             & -162                                        & [2,3,5,6,8,9, \newline{}10,11,12,14,15,18, 19] \\
        \bottomrule
    \end{tabular}%
    \label{tab:detail-solution}%
\end{table}%

Table \ref{tab:detail-solution} shows the results given by the QAOA simulators in different $r$ configurations. At $p=2,3,4,6$, the first possible solution was identified. At $p=5, 100$, the second possible solution was found. As the number of repeated parameterized blocks increased, we observed the increasing time for the quantum simulators to find a solution. In this particular instance, the solution was successfully obtained after only two iterations of the parameterized blocks. Nevertheless, it is anticipated that the number of iterations needed to uncover a solution will likely amplify as the dimensions of the image expand.

QAOA has several limitations. It should be noted that QUBO problems are NP-hard, and the conversion of finding a minimum $s-t$ problem to QUBO does not necessarily make the original problem easier to solve. Furthermore, QAOA is not guaranteed to return correct solutions with finite repetitions of the parameterized blocks. Various attempts to improve the performance of QAOA have been developed. In (\cite{barkoutsosImprovingVariationalQuantum2019}), the authors proposed the Conditional Value-at-Risk as an aggregation function to speed up the classical optimization process by only averaging the best measurements at the read-out stage. The warm starting strategy proposed by Egger and colleagues (\cite{eggerWarmstartingQuantumOptimization2020}) suggests a smarter preparation of initial states by considering a state obtained by a classical procedure.

\section{Conclusion}
We propose and demonstrate the quantum implementation and optimization of a geometric-constraint surface segmentation problem. Future work will include the implementation of the proposed scheme on a real quantum computer and the analysis of its performance on images with varying complexity.


\bibliographystyle{unsrtnat}
\bibliography{references}

\end{document}